\documentclass[aps,pre,preprint,superscriptaddress]{revtex4-1}
\usepackage{times}
\usepackage{mathptmx}
\usepackage{amsmath}
\usepackage{graphicx}
\usepackage{nicefrac}
\usepackage{hyperref}
\usepackage{dcolumn}
\hypersetup{colorlinks=true,urlcolor=blue,linkcolor=blue,citecolor=blue}

\begin{document}


\title{Individual-level evolutions manifest population-level scaling in complex supply networks}


\author{Likwan Cheng}\thanks{Corresponding author. E-mail: lcheng6@ccc.edu}
\affiliation{City Colleges of Chicago, Chicago, Illinois 60601, USA}
\author{Bryan W.\ Karney}
\affiliation{University of Toronto, Toronto, Ontario M5S 1A4, Canada}


\date{9 December, 2018}

\begin{abstract}
Scaling in complex supply networks is a population-level optimization phenomenon thought to arise from the evolutions of the underlying individual networks, but the evolution-scaling connection has not been empirically demonstrated. Here, using individually resolved, temporally serial, and population-scope datasets from public water supply networks, we empirically demonstrate this connection. On the log-log plot, structural properties of individual supply networks trace out evolutionary paths describable as linear projectiles, each characterized by a slope reflecting optimized physical economies of scale and an intercept reflecting morphological adaptation to settlement contexts. The universality in scaling slope coexists with the variability in scaling intercept, so that networks of diverse morphologies advance in time along a ``common evolutionary track". This cross-level observation establishes that individual-level dynamic evolutions cumulatively manifest population-level optimal scaling in complex water supply networks. 

\vskip 24pt
\noindent {\it Citation:} L.\ Cheng and B.\ W.\ Karney, Physical Review E, {\bf 98}, 062323 (2018), published on 28 December 2018, copyright \copyright \ by American Physical Society.

\vskip 3pt
\noindent DOI: \href{https://journals.aps.org/pre/abstract/10.1103/PhysRevE.98.062323}{10.1103/PhysRevE.98.062323}

\end{abstract}

\maketitle
\section{Introduction}
Complex systems---organisms, river-basin landscapes, human settlements, etc.---owe their sustenance to the continuous supplies of essential resources, such as water and energy. The supply network---a web of distributive conduits that connect a central source to all parts of the system---is a complex network and vital infrastructure. A hallmark of the supply network is scaling---the systematic, disproportionate (usually decreased, or allometric) rates of growth in network properties with system size. Scaling is thought to be an evolutionary adaptation by networks to optimize themselves \cite{wes99,rin14}. Indeed, among evolved supply networks, infrastructural allometric scaling has been observed in urban roads \cite{bet13}, drinking water supplies \cite{che17}, and plant leaf venations \cite{sac12}; whereas energy allometric scaling has been observed in river-basin runoffs \cite{rin14,dod10}, water supplies \cite{che17}, and organism metabolic rates as attributed to flows in vascular networks \cite{wes99}. Additionally, dynamical simulations of optimizing supply networks showed the emergence of topological hierarchies, also implying scaling \cite{sin96,hud13,lou13,ron16}.

The origins of scaling as a population optimal phenomenon can be explained from two distinct perspectives. The ``mechanistic" origin, or {\it why}, is explained by theoretical optimization laws; the ``manifest" origin, or {\it how}, is explained by underlying individual evolution dynamics. For the first, to determine population-level optimization, one makes the observation of ``population static (PS) scaling"---the static (instantaneous), internetwork relations of a property among different-sized networks in a population---and compare it to the prediction of the scaling law. For the second, to describe individual-level evolution, one will need to make the different type of observation of ``individual dynamic (ID) scaling"---the temporal, intranetwork relations of a property within an individual network as it grows in size---and compare it to population scaling. In this last step, an agreement between the ID scaling and the PS scaling would indicate that individual dynamics indeed explain population optimization.

Until now, the connection between evolution as an individual-level dynamic process and scaling as a population-level optimal phenomenon has not been empirically demonstrated in any real-world complex supply networks. Previous efforts on network scaling have overwhelmingly focused on the mechanistic origin,  through developing scaling laws and applying them to PS datasets \cite*[e.g.,][]{wes99,ban99,dod10,bet13,che17}. By contrast, works on the manifest origin have been scarce, largely limited by observational challenges in individual dynamics. To our knowledge, only two previous studies have in effect compared ID and PS (or ensemble static) scaling, but both produced results contrary to expectation. One study on the metabolism in 43 species of growing seedling plants finds isometric (linear) metabolic rates with plant mass in ID scaling \cite{rei06}, contrasting the allometric metabolic rates found in a broader range of plants in PS scaling \cite{enq07}. Another study on congestion-induced traffic delays in 101 large American cities finds scaled delay relations with city size in ID scaling, but with different exponents from the ensemble static scaling of the same data \cite{dep18}. Since these studies present properties for only a segment of a population (in the case of seedling plants) or properties not exclusively intrinsic of a network (in the case of traffic delay), the observed disagreements between ID and PS scaling may not be inferable for networks in general. Nonetheless, these unexpected results heighten the need for a direct, unambiguous determination on how ID and PS scaling relate in complex supply networks. 

In this paper, using individually resolved, temporally serial, and population-scope datasets from public water supply (PWS) networks, we simultaneously evaluate the PS and the ID scaling of intrinsic network properties. Using an ``evolutionary track (ET)" model that highlights the coexistence of the universality in scaling slope and the diversity in scaling intercept, we show that individual-level dynamic evolutions cumulatively manifest population-level static scaling. We further use individual-level statistics to explore factors that impact population-level scaling exponents. 

\section{Empirical Results and Analysis}
\subsection{Evolution in scaled networks}\label{sec:evol}
We examine the evolutions of the PWS networks in the U.\,S.\ state of Wisconsin, a population of networks each serving a city, village, town center, or sanitary district; their PS scaling properties were previously reported \cite{che17}. This population of PWSs has the characteristics of a large population size ($n = 582$), a wide size range ($\sim$5 orders of magnitude), and an approximately lognormal size distribution that is characteristic of both organisms and technoeconomic entities \cite{cla09a,mit11}. 

The notion of ``evolution" in the context of scaling requires clarification. First, the meaning of the word needs substantiation. Here, evolution does not mean cross-generation variations; instead, it connotes ``developmental" variations toward increased fitness during growth. In this sense of the word, evolution somewhat resembles development of plant venation networks. Vein development algorithms are quite general \cite{sac13} and to great extent rely on the non-genetic mechanisms of local adaptation and global optimization \cite{ter10,ron16}. For PWSs, local adaptation within an individual network derives from its mutualistic feedback relations with settlement consumers, which evolve the network through local, incremental additions and retirements of water mains---a pattern of infrastructural growth documented in urban histories \cite{smi13}. Global optimization of a network comes from the holistic nature of water supply as a common good of the civil society, governed by public monopolies \cite{bee12} with policies that seek to ``serve every individual and strengthen the whole" \cite{smi13}. 

Second, a distinction needs be made between ``evolution" as change ``{\it to} adapt" versus  ``to {\it renew} adaptedness" during growth \cite{mcd15}. For scaled networks, the word refers to the latter.  In scaling, optimization---or adaptation, or fitness---specifically refers to the attainment of economies of scale as expected by scaling laws. The most basic set of scaling laws describe geometric optimizations of network structures \cite{bet13,che17}. Although actual PWS designs often seek optimization in operations economics instead \cite{kar00,swa08}, the scaling of these two sets of properties are usually mutually consistent \cite{che17}.

If an entity evolves, or adapts itself, from an amorphous to a structured state, the ensued topological change is reflected in a change in the scaling exponent toward the optimal value. Such a transition was in fact reported for settlement scaling in ancient societies over historical times \cite{ort15}. For modern urban networks, one rarely witnesses such transitions since they occur ``virtually" during the design phase, so that in the actual laid infrastructure, adaptedness and optimal scaling are already in place. However, growth would render the network to lose its fitness because property optimality generally scales nonlinearly (allometrically) with size---unless the network renews its adaptedness in step with growth. This renewal again entails changes in topology but not this time in the (already optimized) scaling exponent. Observed temporally, an evolving scaled network in growth is one that marches from adaptedness to adaptedness with a dynamic optimal constancy, rather than change, in the exponent, as the network becomes ever ``scaled up" replicas of its former self. PWS networks are such networks.

\begin{figure}[t]
\centering
\includegraphics[width=0.6\columnwidth]{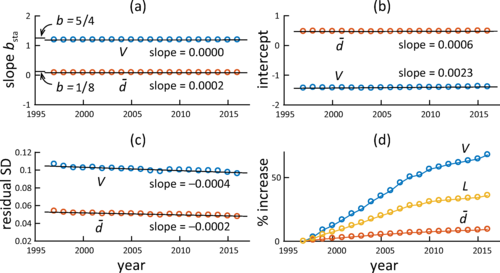} 
\caption{\label{Residual} Population static (PS) scaling. Time series for the scaling of network volume $V$ and mean diameter $\bar d$ with network length $L$ for the population of PWS networks: (a) the scaling exponent $b_{\rm sta}$, (b) the $y$-intercept $y_0$, (c) the residual standard deviation $\sigma$, and (d) the average percentage increases in $V$, $L$, and $\bar d$ of the population. For the raw data and fits that yielded the summary statistic data points in (a-c), see Figs.\ S1-S3 in Supplemental Material (SM) \cite{sm}.}
\end{figure}

\subsection{Scaling}
Specifically, we follow the temporal changes in infrastructural parameters (variables) for each individual network of the population for a 20-year period, from 1997 to 2016. During this period, settlement growths had resulted in average increases in network length $L$, volume $V$, and mean diameter $\bar d$ by 38\%, 65\%, and 9\%, respectively [Fig.\ \ref{Residual}(d)]. The theory-predicted ideal scaling exponents for $L$, $V$, and $\bar d$ with settlement size (as measured by water consumption) are $\frac{2}{3}$, $\frac{5}{6}$, and $\frac{1}{12}$, respectively \cite{che17}. Eliminating size, we establish the scaling of volume $V$ and mean diameter $\bar d$ in relation to length $L$:
\begin{align}\label{eq:laws}
V &= \frac{\pi {\bar{d_0}}^2}{4} L^\frac{5}{4} \\
\bar d &= {\bar d}_0 L^\frac{1}{8},
\end{align}
where the ideal scaling exponents are $\frac{5}{4}$ and $\frac{1}{8}$, respectively, and the scaling prefactors are dependent on the constant mean diameter ${\bar d}_0$. On a log-log plot, the $(x,y)$ coordinate positions of a network variable, where $x= \ln L$, and $y= \ln V$ or $\ln \bar d$, are related by the least-squares (LS) linear model as
\begin{equation}\label{eq:lsm}
y = \hat y + r = b x + y_0 + r
\end{equation}
where $\hat y$ is the model-expected $y$-value and $r$ is data residual, with model slope $b$ (exponent) and intercept $y_0$ (prefactor). 

Let $(x_{ij},y_{ij})$ be the coordinates of individual network $j$ at time $i$ on the log-log plot; $i = 1,2,... 20$ annual times, and $j = 1,2,... 582$ municipal networks. 

\subsubsection{Population static scaling}
PS scaling presents a time-instantaneous (static), population-level view of network relations on the log-log plot. The PS scaling exponent is given by the LS slope:
\begin{equation}\label{eq:b_sta}
b_{\rm sta} =\rho_i (x_{ij}, y_{ij}) \frac{\sigma_i (y_{ij})}{\sigma_i (x_{ij})},
\end{equation}
where $\rho_i$ is the Pearson correlation coefficient and $\sigma_i$ are the variable standard deviations (SD) at a constant time $i$ for all individual networks $j$ of the population. Multiple PS scaling observations over time form a time series. The time-series average of the observed $b_{\rm sta}$ are $1.19\pm 0.01$ for $V$ scaling and $0.08\pm 0.01$ for $\bar d$ scaling, respectively [Fig.\ \ref{Residual}(a)]. The apparent deviations from the ideal values of $\frac{5}{4}$ and $\frac{1}{8}$, respectively, can be attributed to a densification effect on the scaling of $L$ with size. This is due to main duplications that are more prevalent at larger cities and it can be corrected with an empirical exponent correction of $\Delta = 0.03$ in the exponent of $L$ as derived from the fitting of empirical data \cite{che17}. With this correction, the exponent predictions become $\frac{5}{6}/(\frac{2}{3}+\Delta) = 1.196$ for $V$ scaling and $(\frac{1}{12}-\frac{1}{2}\Delta)/(\frac{2}{3}+\Delta) = 0.098$ for $\bar d$ scaling, now agreeable with observations. Notably, the simultaneous temporal occurrence of the disproportionate parameter growths [Fig.\ \ref{Residual}(d)] and the constant exponents [Fig.\ \ref{Residual}(a)] is an evidence of continuous renewals of adaptedness in the population.

\begin{figure}[t]
\centering
\includegraphics[width=0.55\columnwidth]{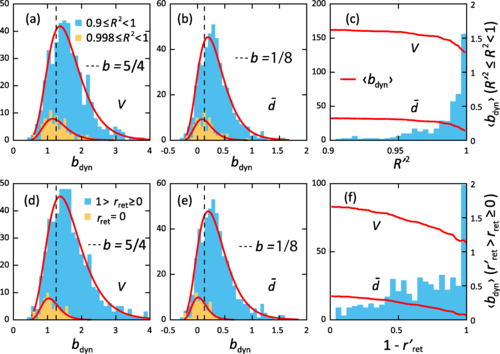}
\caption{\label{IndDist} Individual dynamic (ID) scaling. Distributions of the exponents $b_{\rm dyn}$ for the scaling of (a) $V$ and (b) $\bar d$ with $L$ for projectiles with best-fit adjusted $R^2$ within $0.9 \le R^2 < 1$ and $0.998 \le R^2 < 1$, respectively (histograms), with lognormal distribution fits (solid curves). (c) Number of PWS networks (left axis) and incremental cumulative average (geometric mean) $\langle b_{\rm dyn}\rangle$ for projectiles whose $R^2$ are within $R'^2 \le R^2 < 1$ (right axis), where $R'^2$ ranges from 0.9 to below 1. Distributions of $b_{\rm dyn}$ for the scaling of (d) $V$ and (e) $\bar d$ with $L$ for projectiles with $r_{\rm ret}$ within $1 > r_{\rm ret} \ge 0$ and at $r_{\rm ret}=0 $, respectively, with lognormal distribution fits. (f) Number of PWS networks and incremental cumulative average $\langle b_{\rm dyn}\rangle$ for projectiles whose $r_{\rm ret}$ are within $r'_{\rm ret} > r_{\rm ret} \ge 0$, where $1- r'_{\rm ret}$ ranges from above 0 to 1.}
\end{figure}

\subsubsection{Individual dynamic scaling}
\begin{figure*}[t]
\centering
\includegraphics[width=0.8\textwidth]{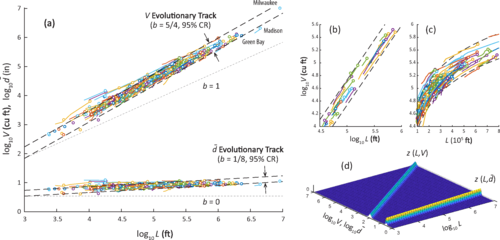} 
\caption{\label{SlopeOne} (a) Each color projectile traces the evolutionary path of the $(L,V)$ or $(L,\bar d)$ coordinates of an individual PWS network in Wisconsin for the 20-year period from 1997 (tail) to 2016 (head, circle) on the log-log plot ($n= 486$). The dashed-lines indicate the 95\% confidence range (CR) of the evolutionary track. (b) Details of the most ``ideal" projectiles, with $|b_{\rm dyn}-b|< 0.07$ for $V$ scaling ($n = 44$). (c) Details of projectile movements on a linear-log plot showing the robustness of directions. (d) Model of the evolutionary track computed with ideal $b$ ($\nicefrac{5}{4}$ for $V$; $\nicefrac{1}{8}$ for $\bar d$) and empirical $\log_{10 }{\bar d}_0$ (0.27) and $\sigma$ (0.10 for $V$; 0.05 for $\bar d$). See time-lapse movie on projectile dynamics in SM \cite{sm}.}
\end{figure*}

ID scaling presents a time-cumulative, individual-level view of network changes on the log-log plot. As each network grows, its coordinate point temporally advances, tracing out an approximately linear ``projectile" [Fig.\ \ref{SlopeOne}(a)]. The slope of the fitted line to the comprising points of the projectile with the LS model is (see Fig.\ \ref{CasePlot} for examples of projectile fits):
\begin{equation}\label{eq:b_dyn}
b_{\rm dyn} =\rho_j (x_{ij}, y_{ij}) \frac{\sigma_j (y_{ij})}{\sigma_j (x_{ij})},
\end{equation}
where $\rho_j$ is the Pearson correlation coefficient and $\sigma_j$ are the variable SD for a given individual network $j$ for all times $i$. We now have a population ensemble of $b_{\rm dyn}$ and $y$-values for $V$ and $\bar d$ scaling, respectively. We consider these statistical distributions below. 

\begin{table}[t]
\caption{\label{table1} Observed exponents for the scaling of $V$ and $\bar d$ with $L$ in population static scaling $b_{\rm sta}$ and individual dynamic scaling $b_{\rm dyn}$, the latter given for specific ranges of projectile linearity $R^2$ and main retirement rates $r_{\rm ret}$.} 
\centering
\begin{ruledtabular}
\begin{tabular}{lllr}     
Scaling exponent\footnote{The $b_{\rm sta}$ values are time-series means and SDs; the $b_{\rm dyn}$ values are population geometric means and 95\% CIs.} 
& \qquad $V$  & \qquad $\bar d$  & $n$\ \ \  \\
\hline 
$b$ (ideal)                  &    \quad\ \ \ $\nicefrac{5}{4}$  &  \quad\ \ \ $\nicefrac{1}{8}$   &   \\
$b_{\rm sta}$            &    $1.19\pm0.01$        &   $0.08\pm0.01$     &   567  \\
$b_{\rm dyn}$\,($0.9\le R^2<1$)   &  $1.53\,(0.79, 2.98)$  &   $0.27\,(-0.10, 1.02)$  & 448  \\
\quad\ \ \,   ($0.998\le R^2<1$)   &  $1.25\,(0.71, 2.19)$  &   $0.12\,(-0.13, 0.57)$  & 83  \\
$b_{\rm dyn}$\,($1> r_{\rm ret}\ge 0$)   &  $1.56\,(0.77, 3.14)$  &   $0.29\,(-0.11, 1.11)$  & 503  \\
\quad\ \ \,   ($ r_{\rm ret}= 0$)  &  $1.10\,(0.68, 1.77)$  &   $0.04\,(-0.17, 0.40)$  & 64  \\
\end{tabular}
\end{ruledtabular}
\end{table}

\begin{figure}[th]
\centering
\includegraphics[width=0.55\columnwidth]{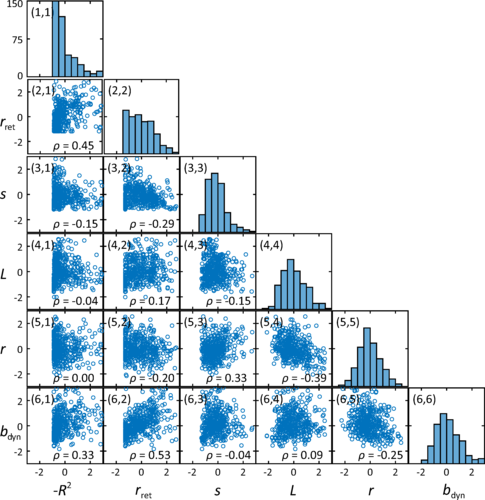}
\caption{\label{Corr} Correlations of individual network parameters. Distribution histograms and pairwise correlation scatterplots for six parameters describing the individual PWS networks: ID scaling exponent $b_{\rm dyn}$, lateral location (residual) $r$, network size (total main length) $L$, network speed of growth $s$, network retirement rate $r_{\rm ret}$, and network projectile linearity $R^2$. Networks included are defined by the range $0.9\le R^2<1$ in Fig.\ \ref{IndDist}(a) for $V$ scaling ($n=448$); the parameters $L$, $s$, $r$ are in log units; the axes of the plots are in standardized units and Pearson correlation coefficients are shown in the plots.} 
\end{figure}

The projectile slopes $b_{\rm dyn}$ show lognormal distributions that are peaked around the law-predicted values of $\frac{5}{4}$ and $\frac{1}{8}$, respectively (Fig.\ \ref{IndDist}). These peaked positions reflect an optimization, or adaptedness, in network geometric (structural) economies of scale, achieved through the formation of hierarchies \cite{bet13,che17}. These geometric economies are associated with the economies in resources, including materials, energy, and operational costs \cite{che17}, considered during network design \cite{kar00,swa08}. Carefully examined, the projectile slopes $b_{\rm dyn}$ show nuanced correlations with two parameters: the linearity of the projectile, described by best-fit adjusted $R^2$; and network main retirement rate $r_{\rm ret}$, defined as the cumulative length ratio of the retired to the added mains in the study period. With increased $R^2$, the cumulative average (geometric mean) of $b_{\rm dyn}$ trends downward to near the ideal $b$. With decreased $r_{\rm ret}$, the cumulative average $b_{\rm dyn}$ also trends downward but slightly passes below the ideal $b$. Since $R^2$ and $r_{\rm ret}$ are partially correlated [Fig.\ \ref{Corr}(2,1)], the dependences of $b_{\rm dyn}$ on both likely rest primarily with $r_{\rm ret}$. The distribution of $b_{\rm dyn}$ contrasts the constant-valued $b_{\rm sta}$, suggesting statistical averaging in the PS approach masks the distribution of individual slopes (Table \ref{table1}).

The projectile ``$y$-intercept" can be equivalently expressed in terms of the individual projectile residual; the projectile residual is defined as the average residual of the projectile's comprising data points, $ r = \langle y_i-\hat y\rangle$, where $y_i$ denotes the $y$-value for year $i$, and $\hat y= bx_i + y_0$, where $y_0$ is the PS $y$-intercept (Eq.\ \ref{eq:lsm}). The distribution of $r$ is approximately normal [Fig.\ \ref{Corr}(5,5)], consistent with the distributions of the PS residuals (Figs.\ S2,\,S3). The peaking of the $r$ distribution around $\hat y$ (standardized $\langle r \rangle = 0.01$) reflects an ``adaptive selection" in network transverse size $\bar d_0$. Unlike physical optimization, this selection is ``ecological" in nature---that is, the choice of the 6- or 8-inch diameters as the ``fittest" terminal main sizes based on domestic and fire-protection demands. The distribution of $r$ represents an ``adaptive divergence"---an evolution of differences as a result of adaptation to different environmental (geographic or civic) conditions \cite{ric14}. In effect, the $r$ distribution quantifies the adaptive morphological diversity of the networks in terms of transverse size.

Each individual network is characterized by six parameters: slope or scaling exponent $b_{\rm dyn}$; residual $r$ (equivalent to intercept); network total length $L$ (a measure of network size); network main retirement rate $r_{\rm ret}$; network projectile linearity $R^2$; and network speed of growth $s$, defined as the head-to-tail distance of the projectile on the log-log plot divided by total lapsed time. The distributions and pairwise correlations of these parameters are shown in Fig.\ \ref{Corr}.

\section{Linking Individual Evolutions and Population Scaling}
\subsection{Common evolutionary track}
We now connect individual-level evolution and population-level scaling by introducing the ``evolutionary track'' model \cite{et}. On the log-log plot, the evolutionary track is defined by a direction based on the ideal $b$; a centerline based on $\hat y= b x + y_0$, where $y_0$ is the PS intercept; and a width based on the SD of the PS residuals $\sigma$. The individual projectiles are then overlain onto the plot, finding themselves residing within the ET and moving in a direction parallel to the track [Fig.\ \ref{SlopeOne}(a-c)]. The angle extended from $b$ to $\langle b_{\rm dyn}\rangle$ (geometric mean) is $4.9^{\circ}$ for the $V$ track. For an idealized, normal density distribution of networks across the track, the probability $z$ of finding a network on the plot is
\begin{equation}\label{eq:model}
z(x, y |b,{\bar d}_0,\sigma) = \frac{1}{\sqrt{2\pi}\sigma}\exp\frac{-(y-\hat y)^2}{2{\sigma }^2},
\end{equation}
where $x=\ln L$ and 
\begin{equation}
y,\, \hat y = \left\{ \begin{array}{l}
				\ln V,\ \ln \big(\dfrac{\pi}{4} {\bar{d_0}}^2 L^\frac{5}{4}\big)  \\
				\ln \bar d,\ \ln({\bar d}_0 L^\frac{1}{8})
			\end{array}
		\right.
\end{equation}
for the $V$ and the $\bar d$ tracks, respectively [Fig.\ \ref{SlopeOne}(d)].

The ET model signifies an integration among ecology (network diversity described by $\sigma$), evolution (selection described by ${\bar d}_0$), and physics (optimization described by $b$), which together parameterize the track (Eq.\ \ref{eq:model}). In this integrative view, the universalness of physical optimization that would have covered the entire $(x,\,y)$ space of the plot is instead truncated by the selection and finite adaptive divergence to a narrow band of ``stability space" (in physics terminology) or ``habitat'' (in ecology terminology) that is the track. This interpretation critically explains the observation: the synchronous motions of the parallel projectiles in which the homogeneity in direction (slope $b$) commanded by universal optimization coexists with the heterogeneity in lateral position (intercepts, or $r$) commanded by network diversity. (The rationale of the ET model is further illustrated through an analogy in Appendix \ref{sec:planets}).

\subsection{Network diversity}
Network diversity, as stated above, is quantified here in terms of network transverse size $\bar d_0$, expressed for the individual as residual $r$ (intercept deviation) and for the population as $\sigma$. These residuals are not random ``errors", but rather are evolved ``character displacements" (in ecology terminology) or ``relaxations" (in physics terminology)---selected deviations from the norm (the ET centerline) as responses to specific environments. In this view, the heterogeneity in intercept is intimately connected to the homogeneity in slope. This heterogeneity-homogeneity interplay is found in many physical and ecological networks in general and in PWS networks in particular, as discussed below. 
 
In dynamic physical networks, it was shown that the homogeneous state of a system may require a heterogeneous system composition \cite{nis16}. Demonstrated with networked oscillators, such a system achieves dynamic stability through the coordination of differences among its constituents (namely, nonidentical oscillators). The stability space of the system possesses a spatial symmetry that excludes compositional homogeneity and within which individual coordinates assume deviated sites. In solid-state (crystalline) networks, especially in minerals, heterogeneity-induced stability is common. For example, in the common rock-forming phillosilicates (layered silicates) of micas, the strains left from the formation of the varied polyhedral layers are released through relaxations of atoms to distorted sites \cite{che03}. This local heterogeneity in atomic sites is a tradeoff made for the global homogeneity in overarching crystal symmetry, and it is the reason for the rich diversity of mica species (end members) found in nature \cite{fer02}. 

In ecological networks, diversity among individuals promotes population stability through competitive exclusion, niche differentiation, and functional complementarity \cite{beg06}.  Diversity among cities translates into diversity in the embedded municipal networks. Geographic conditions play a role in network diversity. Urban networks tend to follow gridiron street patterns with loops and hierarchies, whereas rural networks tend to be expansive and branching. Civic function also plays a role. Industrialization, institutionalization, and commercialization tend to centralize consumptions and widen network transverse sizes, whereas residential communities tend to be geographically disperse and have narrower mains. Settlement functional differentiation---industrial cities, college towns, residential suburbs, and so on---results in a heterogeneous array of network morphologies that follow the specific functions. 

\subsection{Case studies}\label{sec:Case}
Below, referring to Figs.\ \ref{CaseMaps} and \ref{CasePlot}, we illustrate network diversity using selected case networks from the present PWS population. We show that the interplays between scaling slope and intercept (residual) consistently support the arguments of the ET model.

\subsubsection{Franklin and Brookfield: Two young suburban networks} Franklin \cite{fra} and Brookfield \cite{bro} are nearby residential suburbs in the Milwaukee metropolitan area in the populous southeastern region of the state. As young networks, they have very low main retirement rates, $r_{\rm ret}=0.002$ for Franklin and $r_{\rm ret}=0.08$ for Brookfield. Urban sprawl from Milwaukee drives network growth in both cities, but the different lengths of their projectiles, $s = 0.72$ for Franklin and $s = 0.22$ for Brookfield, reflect different speeds of growth speeds. Crucially, a hydrological boundary divides the two communities: Franklin resides within the Great Lakes basin and is entitled to Lake Michigan water but Brookfield straddles just outside of the boundary line and must rely on groundwater. The difference in centrality of the two water sources left distinct traits on the networks' transverse size, separating their (standardized) residuals wide apart on the ET, $r = 0.81$ for Franklin and $r = -1.53$ for Brookfield. Remarkably, their projectile slopes are unaffected by the large difference in residual: $b_{\rm dyn}=1.05$ for Franklin and $b_{\rm dyn}=1.07$ for Brookfield. The Franklin-Brookfield comparison is a striking demonstration of the central argument of the ET model: the coexistence between the universality in slope and the diversity in intercept.

\subsubsection{Ashland: An aging urban network} In contrast to Franklin and Brookfield, Ashland \cite{ash} is an old nineteenth-century industrial city on the rural northwestern shoreline of Lake Superior. With the average main age dates back to 1956 and the city's present population remains unchanged from a century ago, the revitalization of the network has come largely from the replacements of aged and undersized mains, rather than net expansions. This results in a large main retirement rate, $r_{\rm ret} = 0.42$, which in turn leads to a relatively large positive deviation in scaling slope, $b_{\rm dyn} = 2.32$ (see also Sec.\ \ref{sec:life}). On the other hand, the urban centrality and gridiron street patterns of the network give it a relatively centered residual, $r=0.17$. The Ashland network represents an adaptively divergent individual slope, but this is reconcilable with the universal slope for the population in the ET model because such individual divergences are nonsystematic; that is, $b_{\rm dyn}$ lack reasonable correlation with $L$ ($x$-direction) or $r$ ($y$-direction) [Fig.\ \ref{Corr}(6,4),\,(6,5)] (See also \ref{sec:Long}).

\subsubsection{Madison: A network constrained by geography} Madison \cite{mad}, the state's second largest city, has a more balanced civic profile and growth history. But conditions in resource and geography shaped its network into extraordinary morphology. An inland city, Madison is unique among the state's large cities in sourcing groundwater from a distributed network of 23 wells rather than surface water from a centralized treatment plant. Its unusual "Lake, City, Lake" geography---with two large lakes taking up 18\% of its surface area near the city center---pushes consumers to the outskirts. These factors result in a network marked by unusual decentralization and low hierarchy---the largest mains of the Madison network is only 24 inch in diameter, far smaller than typical for its size. This small transverse size results in a residual of $r = -3.02$ (at 0.26 percentile). Even at this extreme deviation, the network's projectile approximately follows the expectation of the scaling law, $b_{\rm dyn}=1.68$. The Madison network showcases the spatial invariance of the scaling exponent, consistent with the spatially universal character of optimization laws assumed in the ET model.

\begin{figure}[t]
\centering
\includegraphics[width=0.55\columnwidth]{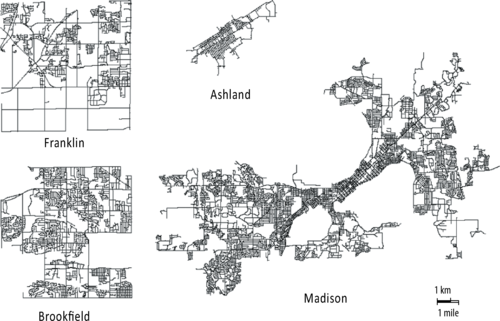}
\caption{\label{CaseMaps} PWS networks of the cities of Franklin \cite{fra}, Brookfield \cite{bro}, Ashland \cite{ash}, and Madison \cite{mad}.} 
\end{figure}

\begin{figure}[t]
\centering
\includegraphics[width=0.4\columnwidth]{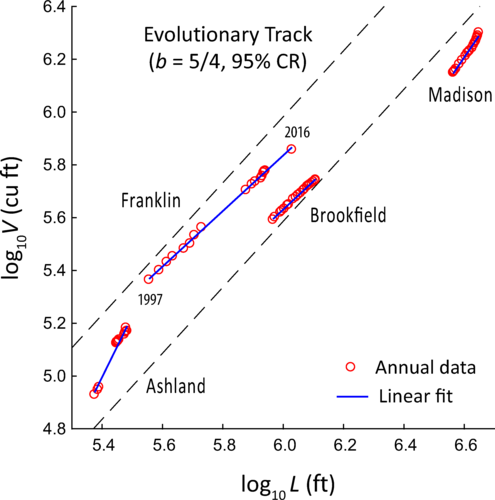}
\caption{\label{CasePlot} Projectiles on the volume-length log-log plot for the PWS networks of Franklin ($b_{\rm dyn}=1.05$), Brookfield ($b_{\rm dyn}=1.07$), Ashland ($b_{\rm dyn} = 2.32$), and Madison ($b_{\rm dyn}=1.68$).} 
\end{figure}

\subsection{Exponent analysis}
ID scaling opens up the capability for statistical analysis of scaling exponents. Here, we explore a few factors that impact scaling exponents. 

\subsubsection{Life history in network scaling}\label{sec:life}
The positive correlation between $b_{\rm dyn}$ and $r_{\rm ret}$, with $\rho = 0.53$ [Figs.\ \ref{Corr}(6,2), S4], can be explained by the fact that a typical main replacement results in an increase in diameter (to accommodate growth) without net changes in length, so that networks with frequent main replacements tend to increase $\bar d_0 $ faster and therefore have higher $b_{\rm dyn}$. By life stage, young settlements tend to be horizontal and its networks have low $r_{\rm ret}$ and low $b_{\rm dyn}$ (less allometric with size) (e.g., Franklin and Brookfield). As settlements age, network mains are widened to accommodate vertical developments \cite{nor71}, resulting in increased $r_{\rm ret}$ and increased $b_{\rm dyn}$ (more allometric with size). This suggests that network topologies follow a ``life history". This view is in line with that for plant venation architectures, in which early- and late-succession species differ in design, with the early design favoring fast growth and the late design better suited for conduit efficiency \cite{mcc11}. Furthermore, the trends of $b_{\rm dyn}$ with $r_{\rm ret}$ seen in PWS networks (Table \ref{table1}) appear to relate to similar trends seen in plant metabolism, in which small seedlings tend to scale isometrically \cite{rei06} while large trees tend to scale allometrically \cite{enq07}. For PWS networks, since $r_{\rm ret}$ does not appreciably correlate with $r$ ($y$-direction) or $L$ ($x$-direction) [Fig.\ \ref{Corr}(5,2),\,(4,2)], the dependence of $b_{\rm dyn}$ on $r_{\rm ret}$ does not produce distortions in the scaling slope.

\begin{figure}[b]
\centering
\includegraphics[width=0.55\columnwidth]{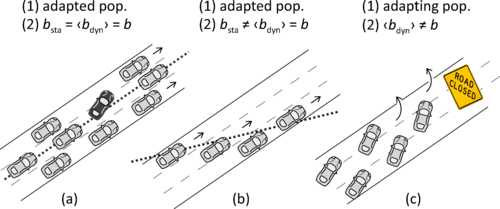}
\caption{\label{cars} ``Cars-on-road" analogy of scaling slopes. (a) In an adapted population with symmetric lateral distribution on the evolutionary track, both $b_{\rm sta}$ (dotted line) and $\langle b_{\rm dyn}\rangle$ (arrows) align with the ideal $b$, even though there may be occasional deviating individuals. (b) In an adapted population with asymmetric lateral distribution, $b_{\rm sta}$ fails to follow $\langle b_{\rm dyn}\rangle$. (c) If the population is actively adapting to a changing circumstance (road closure), the slope and intercept of the existing scaling law break down.} 
\end{figure}

\subsubsection{Long-term change of the evolutionary track}\label{sec:Long}
The ET is a constant in the time scale of projectile movements, as justified by the parameter steadiness in the PS time series (Fig.\ \ref{Residual}). However, in the long term, small biases of $b_{\rm dyn}$ with $r$ could gradually modify the track. We note that $b_{\rm dyn}$ correlates very weakly, negatively with $r$, with $\rho = -0.25$ [Fig.\ \ref{Corr}(6,5)], suggesting that projectiles below the track centerline have slightly higher slopes than those above. This small lateral bias in $b_{\rm dyn}$ may explain the minimally decreasing SD observed in the PS time series [Fig.\ \ref{Residual}(c)], and suggests an eventual narrowing of the track. This implies a long-term evolutionary convergence in network transverse width distribution. Interestingly, a long-term evolutionary divergence was observed in mammalian body mass distribution at the geologic time scale \cite{cla09b}.

\subsubsection{Deviations in population static exponents}
By definition, the PS exponent $b_{\rm sta}$ represents a positional association of data points on the log-log plot. Its inference for dynamic evolution is valid only when the residual distribution is sufficiently symmetric, as illustrated in an analogy in Fig.\ \ref{cars}(a); otherwise, artificial deviations may result. In the present data, a mild local downward asymmetry in the $r$ distribution appears at large $L$, where data become scarce, giving an overall $\rho = -0.39$ [Fig.\ \ref{Corr}(5,4)]. This distortion renders $b_{\rm sta}$ to no longer precisely follow $b_{\rm dyn}$. This local asymmetry is likely partly responsible for the slightly lower observed $b_{\rm sta}=1.19$ compared to the ideal $b=\frac{5}{4}$ [Fig.\ \ref{Residual}(a); Table\ \ref{table1}]. Data distribution distortions could be one reason why observed exponents in PS scaling often differ from predictions. In organism metabolic scaling, the expectation of universal exponents \cite*[e.g.,][]{wes99} was often contradicted by observed deviated slopes \cite{whi10} or curvatures \cite{kol10}. The present work helps reconcile such contradictions by revealing one potential underlying cause: $b_{\rm sta}$ as a measure of ensemble positional association does not always follow $\langle b_{\rm dyn}\rangle$ as a measure of individual evolutionary directions [Fig.\ \ref{cars}(b)]. It follows that deviations in the static exponent are not a strong criterion for refuting scaling laws. The present work supports the notion of universal exponents because $b_{\rm dyn}$ shows little correlation with either speed $s$, with $\rho = -0.04$ [Fig.\ \ref{Corr}(6,3)], or size $L$ ($x$-direction), with $\rho = 0.09$ [Fig.\ \ref{Corr}(6,4)], consistent with a geometric nature of scaling---except that this universality lies basically with $b_{\rm dyn}$ and only conditionally with $b_{\rm sta}$.

\subsubsection{Actively adapting properties}
While intrinsic network properties such as $V$ and $\bar d$ are stable at the projectile time scale, properties at least partially extrinsic of the adapted network may be susceptible to external influences and undertake new adaptations, resulting in changing scaling intercept and slope [Fig.\ \ref{cars}(c)]. This may be said of many complex urban properties, which are often influenced by multiple, sometimes poorly defined factors besides the network itself \cite{cot17}. In particular, this may explain why traffic delay as an urban phenomenon does not show agreements between individual dynamics and ensemble scaling \cite{dep18}, in contrast to the structural properties of PWS networks. Thus, in relating individual dynamics and population scaling, extrinsic urban properties undergoing active ``adaptation" should not be expected to behave the same as intrinsic network properties in a state of continuous ``adaptedness" (Sec.\ \ref{sec:evol}).

\section{Conclusion}
This work demonstrates that individual-level dynamic evolutions explain population-level optimal scaling in complex supply networks. It fulfills an often-assumed, occasionally questioned, but never before demonstrated key expectation in network scaling. This was enabled by the individual dynamic scaling approach, and the individually resolved, temporally serial, and population-scope datasets. The evolutionary track model that connects individual evolutions and population scaling is highlighted by the coexistence of a heterogeneity in intercept (related to network morphology) and a homogeneity in slope. The model's underlying ecology-physics-evolution integrative view represents an extension from the eco-evolutionary integrative view already established in evolution research \cite{toj17}. The individual dynamic scaling data open up statistical analysis for scaling exponents, allowing quantitative explorations in various topics, such as the causes for deviation in scaling exponents. The results presented here should be applicable to other types of complex supply networks, both anthropogenic and natural.

Water supply networks are a hierarchically ordered ``network of networks", given its well-defined population size distribution and property scaling. The present work represents a trans-scale endeavor in property scaling from individuals to the population. To further this end, the multiscale approach that allows unfolding networks across scales based on scaling in probabilistic structural distributions \cite{pez18} may be extended to scaling in deterministic properties. Furthermore, as embedded urban infrastructures, PWS networks could also shed light on cross-scale relations in a ``system of cities" \cite{bat13}. For water supply networks specifically, this work increases the resolution of scaling from the population (regional policy) level to the individual (municipal operations) level. The predictabilities entailed from these scaling relations have wide-ranging practical applications in the design and management of water and other vital civic infrastructures, as well as that of the human settlements that mutualistically embed these networks.

\begin{acknowledgments}
We thank Bruce Schmidt for data explanation, Adilson Motter for beneficial discussions, and two anonymous referees for constructive reviews. 
\end{acknowledgments}

L.C.\ conceptualized the research and contributed expertise in network physics; B.W.K.\ contributed expertise in water distribution systems engineering.

\begin{appendix}
\section{``Planetary gravity" analogy of the evolutionary track model}\label{sec:planets}
\begin{figure}[t]
\centering
\includegraphics[width=0.55\columnwidth]{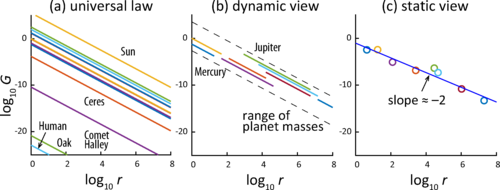} 
\caption{\label{Planets} ``Planetary gravity" analogy of the evolutionary track model.}
\end{figure}

The gravitational potential $G$ of a mass $M$ at a distance $r$ is described by Newton's law of gravitation $G\propto M/r^2$, which can be written in log-log space as $\log G \propto \log M -2\log r$. For any constant mass $M$, a line of slope $-2$ relates $G$ and $r$ in the log-log plot. This universal law applies to all masses, so that parallel lines drawn for different $M$ fill the entire $(r,\,G)$ space (Fig.\ \ref{Planets}a). Planets are ``selected" masses bound in size between dwarf planets and dwarf stars; this selection includes only a narrow band of the general $(r,\,G)$ space. As planets move in their orbits, their distances $r$ from an arbitrary observer vary, resulting in observed dynamic traces of lines obeying the law (Fig.\ \ref{Planets}b). If the same observation is made for only an instantaneous moment of time (a snapshot), the planets are ``frozen" at some momentary points. A linear fit to these static points recovers the slope $\sim -2$, but it does not reveal the underlying dynamic traces (Fig.\ \ref{Planets}c). Fig.\ \ref{Planets}(b) appears like the ID scaling in Fig.\ \ref{SlopeOne}(a); Fig.\ \ref{Planets}(c) appears like the PS scaling in Fig.\ S1. 

\section{Data and methods}\label{sec:data}
Data on public water supply networks used in this study drew from the annual reports of public water utilities obtained from the Wisconsin Public Service Commission, the state regulatory authority \cite{wis13}. The general procedures of data extraction were described previously \cite{che17}. There are 582 operating PWSs; this number varies slightly over time. The processed final overall dataset is a three-dimensional matrix. The dimensional parameters and their sizes are: network parameter $p$ ($L$, $V$, $\bar d$; 3 data columns), year $y$ (20 first-of-the-year and 20 end-of-the-year data columns), and water supply utility $n$ (582 data columns). This gives the number of comprising cells: $\sum_1^3 p \sum_1^{20\times 2} y \sum_1^{582} n = 698\,400$.

For the time series analyses of PS scaling, data screening was performed to eliminate only utilities that were both statistical outliers and were in operation for only parts of the study period, so that their inclusion would artificially distort ordinary distributions of the time series from one year to the next. The final average number of PWSs is $n=567$. 

For the ensemble analysis of ID scaling, the following data screening was performed to exclude data lacking linearity and data whose linear slopes are statistical outliers. First, networks without growth in the study period (singularities) were excluded. Second, network projectiles were fitted for an LS linear slope as individual dynamic exponents $b_{\rm dyn}$. Third, the $b_{\rm dyn}$ exponents obtained above were then fitted with a lognormal distribution, which identified their ranges as $0 \le b_{\rm dyn}\le 4$ for $V$ scaling and $-0.5 \le b_{\rm dyn} \le 2$ for $\bar d$ scaling; networks outside these respective ranges were considered as statistical outliers and excluded. Of the remaining networks, networks that meet the linearity criterion of having adjusted $R^2$ values within the range $0.9\le r^2<1$ according to the LS fits for $V$ scaling were included for formal analyses. The dataset passing this final screening comprises $n=448$ networks. When the no-growth (singularity) networks are included, the total number of networks is $n=486$.

The correlation scatterplots shown in Fig.\ \ref{Corr} are based on data with linearity of adjusted $R^2$ within the range $0.9\le R^2<1$ for the $V$ scaling, with $n = 448$. Correlations were also examined for data with network main retirement rate $r_{\rm ret}$ within the range $0> r_{\rm ret}\ge 0$, with $n=503$. These analyses resulted in consistent correlation coefficients.
\end{appendix}

\end{document}